\documentclass[manuscript]{aastex}
\usepackage{emulateapj5, apjfonts}
\epsscale{1.6}
\slugcomment{Accepted for publication in ApJ}
\newcommand{\nuc}[2]{${}^{#2} \rm #1$}

\def\gtaprx {\lower .14ex\hbox{\rlap{\raise .9ex\hbox{\hskip .3ex
	{\ifmmode{\scriptscriptstyle >}\else
		{$\scriptscriptstyle >$}\fi}}}
	\kern -.4ex{\ifmmode{\scriptscriptstyle \sim}\else
		{$\scriptscriptstyle\sim$}\fi}}}
\def\ltaprx {\lower .14ex\hbox{\rlap{\raise .9ex\hbox{\hskip .3ex
	{\ifmmode{\scriptscriptstyle <}\else
		{$\scriptscriptstyle <$}\fi}}}
	\kern -.4ex{\ifmmode{\scriptscriptstyle \sim}\else
		{$\scriptscriptstyle\sim$}\fi}}}

\newcommand{\rr}{{\it r}}
\newcommand{\pp}{{\it p}}

\newcommand{\mev}{\, {\rm MeV} }

\newcommand{\cm}{\, {\rm cm} }
\newcommand{\s}{ \, {\rm s} }
\newcommand{\K}{ {\,\rm K} }

\newcommand{\erg}{\, {\rm erg} }
\newcommand{\ergs}{\, {\rm ergs} }

\newcommand{\km}{{\, \rm km}}

\newcommand{\psec}{\,{\rm s}^{-1}}
\newcommand{\gpccm}{{\, \rm g\,cm^{-3}}}

\newcommand{\del}[2]%
{\frac{\mathrm{d}{#2}}{\mathrm{d}{#1}}}
\newcommand{\Del}[2]%
{\frac{\mathrm{D}{#2}}{\mathrm{D}{#1}}}
\newcommand{\ddel}[2]%
{\frac{\mathrm{d}^2{#2}}{\mathrm{d}{#1}^2}}

\newcommand{\pdel}[2]%
{\frac{\partial{#2}}{\partial{#1}}}
\newcommand{\pddel}[2]%
{\frac{\partial^2{#2}}{\partial{#1}^2}}

\renewcommand{\vec}[1]{\mathbf{#1}}

\newcommand{\Ms}{M_{\odot}}
\newcommand{\DMs}{M_{\odot}\rm \,s^{-1}}

\shorttitle{Nucleosynthesis in collapsar jets}
\shortauthors{Fujimoto et al.}

\begin{document}

\title{Nucleosynthesis in Magnetically Driven Jets from Collapsars}

\author{
Shin-ichiro Fujimoto\altaffilmark{1},
Nobuya Nishimura\altaffilmark{2}, 
and Masa-aki Hashimoto\altaffilmark{2},
}

\altaffiltext{1}{
Department of Electronic Control, 
Kumamoto National College of Technology, 
%%2659-2 Suya, Goshi, 
Kumamoto 861-1102, Japan;
fujimoto@ec.knct.ac.jp.}

\altaffiltext{2}{
Department of Physics, School of Sciences, 
Kyushu University, Fukuoka 810-8560, Japan.}

%%%%%%%%%%%%%%% Author

%%%%%%%%%%%%%%% Abstract
\begin{abstract}
We have made detailed calculations of the composition of magnetically driven jets 
ejected from collapsars, or rapidly rotating massive stars, 
based on long-term magnetohydrodynamic simulations of their core collapse  with various distributions of magnetic field and angular momentum before collapse.
We follow the evolution of the abundances of about 4000 nuclides from the collapse phase 
to the ejection phase and through the jet generation phase using a large nuclear reaction network. 
We find that the \rr-process successfully operates only in energetic jets ($> 10^{51}\ergs$), 
such that U and Th are synthesized abundantly, 
even when the collapsar has a relatively weak magnetic field ($10^{10}\,\rm G$) 
and a moderately rotating core before the collapse. 
The abundance patterns inside the jets are similar to those of the \rr-elements in the solar system. 
About 0.01-0.06 $\Ms$ neutron-rich, heavy nuclei are ejected from a collapsar
with energetic jets.
The higher energy jets have larger amounts of \nuc{Ni}{56}, varying from 3.7$\times 10^{-4}$ to 0.06$\Ms$.
Less energetic jets, which eject small amounts of \nuc{Ni}{56},
could induce a gamma-ray burst (GRB) a supernova, such as GRB 060505 or GRB 060614.
Considerable amounts of \rr-elements are likely to be ejected from GRBs with hypernovae, 
if both the GRB and hypernova are induced by jets that are driven near the black hole.
\end{abstract}
%%%%%%%%%%%%%%% Abstract

%%%%%%%%%%%%%%% Keywords

%%%%%%%%%%%%%%% Keywords

%%%%%%%%%%%%%%% Section 1
\section{Introduction}

%%%%% for a correct counter of footnote
%%\setcounter{footnote}{0}

%%%%%%%%%%%%%%%%%%% Collapsar for R-process sites
During the collapse of a star more massive than 35-$40\Ms$,
the stellar core is considered to promptly collapse to a black hole~\citep{heger03}.
If the star has sufficiently high angular momentum before collapse, 
an accretion disk forms around the hole, and
jets has been shown to be launched from the inner region of the disk
through magnetic processes~\citep{proga03,mizuno04a,mizuno04b,fujimoto06}
and neutrino heating~\citep{nagataki07}.
Gamma-ray bursts (GRBs) are expected to be driven by the jets.
This scenario for GRBs is referred to as the collapsar model~\citep{w93}.
Assisted by accumulating observations that imply an association between 
GRBs and the deaths of massive stars~\citep[e.g.,][]{galama98,hjorth03,zkh04}, 
this model seems to be most promising.

%%%%%%%%%%%%%%%%%%% Nucleosynthesis in Collapsar \& GRB jets
For accretion rates greater than $0.1\DMs$,
the accretion disk is so dense and hot 
that nuclear burning is expected to proceed efficiently.
In fact, the innermost region of the disk, thought to be related to GRBs, 
becomes neutron-rich through electron capture
on protons~\citep{PWH03, fujimoto04, fujimoto05a}.
Nucleosynthesis inside the outflows from the neutron-enriched disk 
has been investigated with steady, one-dimensional models of the disk 
and the outflows~\citep{pruet04, fujimoto04, fujimoto05b}.
Not only neutron-rich nuclei~\citep{fujimoto04}, 
but also \pp-nuclei~\citep{pruet04,fujimoto05b} has been shown to be produced
inside the outflows.
This co-production of \rr- and \pp-elements in the outflows
has been confirmed with a more elaborate calculation of the nucleosynthesis
in jets from a collapsar~\citep{fujimoto07},
based on two-dimensional magnetohydrodynamic (MHD) simulations~\citep{fujimoto06};
the calculation, however, was performed only for a collapsar with 
a strong magnetic field ($10^{12}\rm\,G$) and 
rapidly rotating core, with angular velocity of $10\rm\,rad\,s^{-1}$.
%%%%%%%%%%%%%%%%%%% Present Paper
In the present study, we recalculate the chemical composition of jets from
collapsars, which are shown to eject magnetically driven jets~\citep{fujimoto06},
to investigate the dependence of jet composition on 
the collapsar rotation rate and magnetic field.

%%%%%%%%%%%%%%%%%%% Ni56, GRB without SN
Moreover, 
the recent discovery of nearby GRBs (060505 and 060614)~\citep{fynbo06,gehrels06} 
that show no association with a supernova (SN) 
has led to the suggestion 
that these GRBs were not produced by a collapsar~\citep{galyam06,della06,zhang07}.
The absence of a SN reveals that very small amounts of \nuc{Ni}{56} ($< 0.001\Ms$) 
were ejected from these GRBs~\citep{fynbo06,gehrels06}, in contrast to the cases of GRBs with SNe; 
the SNe associated with GRB 980425 and GRB 030329, SN 1998bw and SN 2003dh, respectively, 
emitted large amounts of \nuc{Ni}{56} ($>0.3\Ms$)~\citep{iwamoto98,maeda03}.
We show here that the masses of \nuc{Ni}{56} ejected from collapsars 
are different from one to another, and as such, the diversity in the amount of \nuc{Ni}{56} 
is unlikely to contradict the collapsar model.

In \S 2, we briefly describe 
the numerical code used for the MHD calculation of the collapsars, 
or rapidly rotating massive stars, 
the initial conditions of the stars before core collapse, 
and the properties of the jets that are emitted.
In \S 3, we present the Lagrangian evolution of the ejecta through the jets and 
the large nuclear reaction network used to follow the abundance evolution of the jets.
It is also shown that the \rr-process operates inside the jets, 
even when the magnetic field is $10^{10}\rm\,G$ and 
the core is moderately rotating, with an angular velocity of $2.5\rm\,rad\,s^{-1}$.
The masses of \nuc{Ni}{56} and \rr-elements ejected from the collapsars
are shown as a function of the energy of the ejecta.
We discuss the effects of neutrino interactions and the energy, 
liberated through nuclear reactions on nucleosynthesis in the jets 
and the dynamics of the collapsars in \S 4. 
Finally, we summarize our results in \S 5.

%%%%%%%%%%%%%%% Section 2
\section{MHD Calculations of Collapsars}\label{sec:mhd}

We have carried out two-dimensional, Newtonian MHD calculations of 
a collapse of rotating, massive (40$\Ms$) star
whose core is assumed to promptly collapse to a black hole.
Here we present our numerical models and the results of the MHD simulations, 
in particular, the production and properties of the jets, which are important for nucleosynthesis.
For details of the MHD models and results, the reader is referred to \citet{fujimoto06}.

\subsection{Input Physics and Numerical Code}\label{sec:model}

The numerical code for the MHD calculations
employed in this paper is based on the ZEUS-2D code~\citep{sn92}
and is the same as used in \citet{fujimoto07}.
We have extended the code to include
a realistic equation of state (EOS)~\citep{kotake04} based on 
relativistic mean field theory \citep{shen98}.
For the lower density regime ($\rho < 10^5 \gpccm $), 
where no data are available in the Shen EOS table, 
we use another EOS~\citep{bdn96}.
We consider neutrino cooling processes.
The total neutrino cooling rate is evaluated with 
a simplified neutrino transfer model based on 
the two-stream approximation~\citep{dpn02}, 
with which we can treat the optically thin and thick regimes 
of the neutrino reactions approximately.
We ignore resistive heating, the properties of which are highly uncertain~\citep{proga03}.
We note that viscous heating is not taken into accounts.
We assume the fluid is axisymmetric and has mirror symmetry about the equatorial plane.
Spherical coordinates, ($r, \theta, \phi $) are used in our simulations and the computational domain 
extends over $50 \km \le r \le$ 10,000$\km$ and $0 \le \theta \le \pi/2$
and is covered with $200(r) \times 24 (\theta)$ meshes.
Fluid is freely absorbed through the inner boundary at 50km, 
which mimics the surface of the black hole. The mass of the black hole is 
continuously increased by the infall of gas through the inner boundary.
We mimic strong gravity around the black hole in terms of a pseudo-Newtonian potential~\citep{pw80}.

\subsection{Initial Conditions}\label{sec:init-cond}

We set the initial profiles of the density, temperature, and electron fraction 
to those of the \citet{hashimoto95} spherical model of a 40$\Ms$ massive star 
before the collapse.
The radial and azimuthal velocities are set to zero initially
and increase as a result of the collapse induced by the central hole and 
self-gravity of the star.
The computational domain extends from the Fe core to an inner
O-rich layer and encompasses about 4$\Ms$ of the star.
The boundaries of the Si-rich layer between 
the Fe core and the O-rich layer are located 
at about 1800 km (1.88$\Ms$) and 3900km (2.4$\Ms$), respectively.
We adopt an analytical form for the angular velocity 
$\Omega$ of the star before the collapse:
\begin{equation}
 \Omega(r) = \Omega_0 \frac{R_0^2}{r^2 + R_0^2},
 \label{eq:omega0}
\end{equation}
as in previous studies
of collapsars~\citep{mizuno04a, mizuno04b} and SNe~\citep{kotake04}.
Here $\Omega_0$ and $R_0$ are parameters of our model.
We consider three sets of ($\Omega_0, \, R_0$): 
($10\,\rm rad\psec$, 1000km) (rapidly rotating core) 
($2.5\,\rm rad\psec$, 2000km) (moderately rotating core)
and 
($0.5\,\rm rad\psec$, 5000km) (slowly rotating core).
We note that for these values of $\Omega_0$ and $R_0$, 
the maximum specific angular momentum is about $10^{17}\,\cm^2\psec$, 
which is comparable to that of the Keplerian motion at 50km
around a 3$\Ms$ black hole.
Therefore, the centrifugal force can be larger than 
the gravitational force of the central black hole, 
and the formation of a disk like structure is expected near the hole.
The initial magnetic field is assumed to be uniform and 
parallel to the rotational axis of the star.
We consider cases with initial magnetic fields 
($B_0$) of $10^8, 10^{10}$, and $10^{12}$G.
It should be noted that 
the magnetic pressure is much smaller than the gas pressure initially,
even for $B_0 = 10^{12}$G.
We have performed MHD simulations for nine models, 
labeled by R8, R10, R12, M8, M10, M12, S8, S10, and S12, 
where the first character, R (rapidly rotating core), 
M (moderately rotating core),
or S (slowly rotating core), indicates the set of $\Omega_0$ and $R_0$
and the numeral, 8, 10, or 12, equals $\log B_0(\rm G)$.
Note that the MHD simulations for collapsars with a moderately rotating core
(M8, M10, and M12) are in addition to those presented in \citet{fujimoto06}.

\subsection{Properties of Jets}\label{sec:jets}

We briefly describe the results of our MHD calculations~\citep{fujimoto06}.
The Lagrangian evolution of the physical quantities inside the jets, 
which is important for nucleosynthesis, is described in \S 3.1 in detail.

We find that jets are magnetically launched from the central region 
of the star along the rotational axis;
after material reaches the black hole with a high angular momentum
of $\sim 10^{17} \rm cm^2 \psec$, 
a disk is formed inside a surface of weak shock, 
which appears near the hole and propagates outward slowly, 
because of the centrifugal force.
The magnetic fields are chiefly amplified by the wrapping of 
the field inside the disk, so that 
the toroidal component increases and dominates over the poloidal component.
Eventually, the jets can be driven by the tangled magnetic field lines
at the polar region near the hole. %% at $t = 0.20 \s$.

We find that jets are launched for six models, 
R10, M10, S10, R12, M12, and S12.
Their properties are summarized in Table 1.
Columns (2) through (4) list the initial conditions for the models.
Column (5) gives the time at the end of each run, $t = t_f$, 
after the onset of collapse, or $t = 0$.
Column (6) gives $t_{\rm jet}$, the time when the jets pass through 1000km.
Columns (7), (8), (9), and, (10) represent 
the mass and magnetic, kinetic, and internal energies of the jets,
respectively. 
We note that the jets are more energetic for 
the collapsars with a moderately rotating core (M10 and M12) than for R12.
Larger amounts of gravitational energy are liberated in 
the central parts of collapsars M10 and M12 than in R12.
This is because the centrifugal force of the core for R12 is
larger than those for M10 and M12.
The accretion rates onto the black hole through the accretion disk 
are larger for M10 and M12 than for R12;
the rates are $\sim 0.1 \Ms\psec$ for M10 and M12 and $\sim 0.01 \Ms\psec$ for R12 
at the epoch of the launching of the jets.

%%%%%%%%%%%%%%% Section 3
\section{Nucleosynthesis in jets}\label{sec:nucleosynthesis}

We next examine nucleosynthesis in jets from collapsars, 
models R10, M10, S10, R12, M12, and S12, based on the results
of the MHD simulations~\citep{fujimoto06}.
We first describe the Lagrangian evolution of the ejecta through the jets in detail
and then proceed to nucleosynthesis inside the jets.

%%%%%%%%%%%%%%%%%%%%%%%%%%%%%%%%%%%%%%%%%%%%%%%%%%%
%% Figure: maximum densities and temperatures
%%%%%%%%%%%%%%%%%%%%%%%%%%%%%%%%%%%%%%%%%%%%%%%%%%%
\begin{figure*}[ht]
 \plottwo{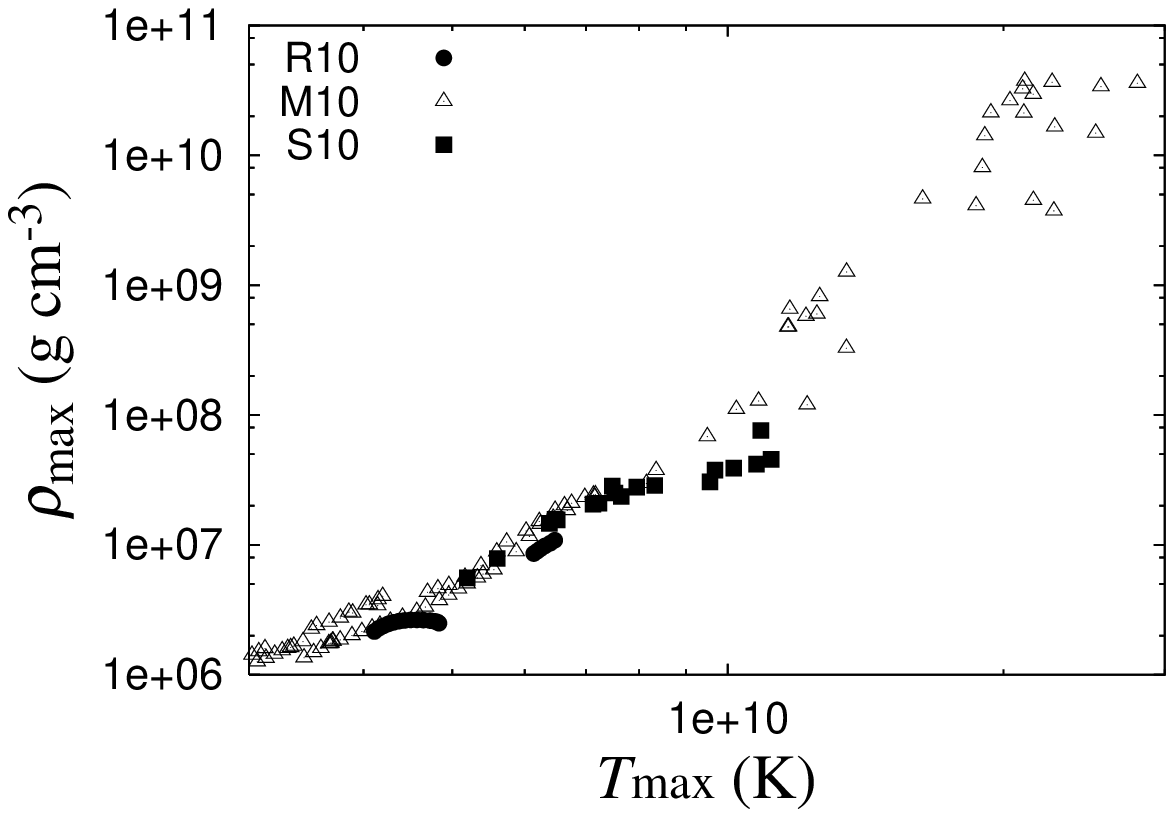}{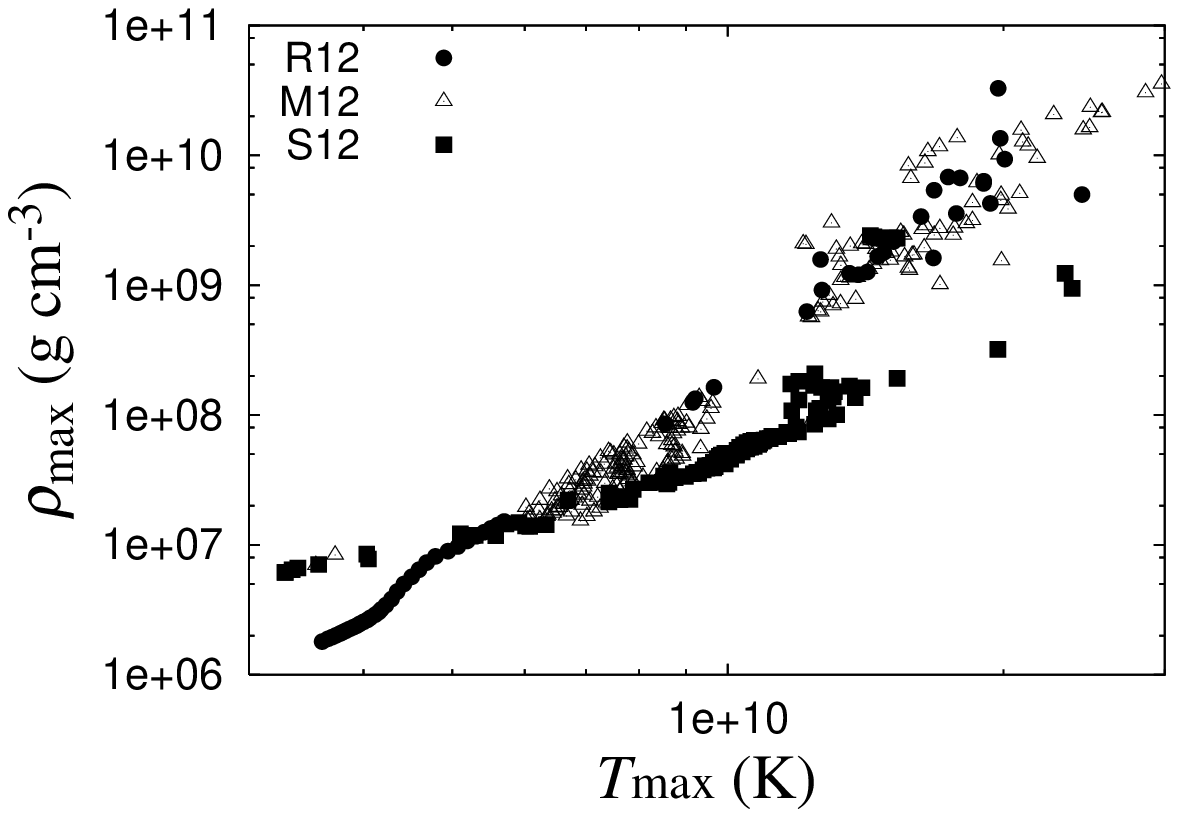}
\caption{
Maximum densities, $\rho_{\rm max}$, and temperatures, $T_{\rm max}$, of jet particles
for models with $B_0 = 10^{10}\, \rm G$ (left)
and $B_0 = 10^{12}\, \rm G$  (right).
The circles, triangles, and squares indicate
models with rapidly, moderately, and slowly rotating cores, respectively.
} \label{fig:peak}
\end{figure*} 

%%%%%%%%%%%%%%% Section 3.1
\subsection{Lagrangian Evolution of Ejecta through Jets}\label{sec:ejecta}

In order to calculate the chemical composition of the jets, 
we need the Lagrangian evolution of physical quantities, such as 
the density, temperature, and velocity of the jet material.
We adopt a tracer-particle method~\citep{nagataki97}, 
as in \citet{fujimoto07}, 
to calculate the Lagrangian evolution 
from the Eulerian evolution obtained from our MHD calculations.
Particles are initially placed between the Fe core and an inner O-rich layer. 
The total numbers of the particles $N_p$ are set to be 
50,000, 1000, 2000, 1000, 2000, and 5000
for R10, M10, S10, R12, M12, and S12, respectively, with which 
we can follow the ejecta through the jets appropriately.
We show how different values of $N_p$ change the masses and abundances of the ejecta through the jets in \S \ref{sec:composition}.
The number of particles in a layer are weighted according to the layer's mass.
We find that 21, 105, 18, 59, 214, and 113 particles are ejected through the jets 
for models R10, M10, S10, R12, M12, and S12, respectively.
Hereafter, we refer to these as {\itshape jet particles}.

%%% adiabatic expansion
We continued the MHD calculations until $t = t_f$ (Table 1).
After $t_f$, we assume that the particles are adiabatic and 
expand spherically and freely.
Therefore, the velocity, position, density, and temperature of each particle are
set to be $v(t) = v_0$, $r(t) = r_f +v(t)(t -t_f)$, 
$\rho(t) = \rho_f(r_f/r(t))^3$, $T(t) = T_f(r_f/r(t))$, respectively, 
where $v_0$ is constant in time and set to $v_f$.
Here $v_f$, $r_f$, $\rho_f$, and $T_f$ are 
the velocity, position, density, and temperature of the particle at $t_f$,
respectively.
It should be noted that in \citet{fujimoto07}
we discussed how the assumption regarding the expansion affects 
the abundance changes in the expansion phase of the ejecta for a low $Y_e$ particle
in model R12, and we concluded that the particle composition depends only weakly 
on this assumption.
Here, $Y_e$ is the electron fraction at $9 \times 10^9 \,\rm K$.

%%% Trajectories
For the strongly magnetized collapsars (R12, M12, and S12), 
the jet particles are initially located inside the Fe core
or just above it, as shown in Figure 1 of \citet{fujimoto07} for R12, 
while the particles are initially placed in the O- or Si-rich layers 
for the mildly magnetized collapsars (R10, M10, and S10).
As the particles fall toward the black hole, their densities and temperatures
rise, and then they decrease when the particles are ejected through the jets, 
as shown in Figure 2 of \citet{fujimoto07} for R12.
Figure \ref{fig:peak} shows 
the maximum densities and temperatures of the jet particles, $\rho_{\rm max}$ and $T_{\rm max}$.
Particles with relatively low density ($< 10^9 \gpccm$) 
have $\rho_{\rm max}$ and $T_{\rm max}$ similar to 
those of ejecta from Type II SNe~\citep{thielemann98}.
On the other hand, for M10, R12, and M12, 
some jet particles have extremely high densities and temperatures 
($\rho_{\rm max} \ge 10^{9} \gpccm$ and $T_{\rm max} \ge 10^{10} \K$), 
which cannot be realized in the ejecta from a Type II SN.
These particles are dense and hot
enough for electrons to be captured on protons, so that the particles become neutron-rich,
as we show below (Figure \ref{fig:peak-ye}).
Hence, the \rr-process is expected to operate in the jets for M10 and M12, 
as well as R12, in which the \rr-process has already been shown to 
operate successfully~\citep{fujimoto07}.

%%%%%%%%%%%%%%% Section 3.2
\subsection{Nuclear Reaction Network}\label{sec:network}

As shown in Figure \ref{fig:peak}, 
some ejecta can attain to temperatures higher than 
$9 \times 10^9 \K$ near the black hole, 
at which point the material is in nuclear statistical equilibrium (NSE).
The abundances of material in NSE
have simple analytical expressions~\citep{clayton68}, 
which are specified by $\rho$, $T$, 
and the electron fraction, as in \citet{fujimoto05a}.
The electron fraction changes through 
electron and positron captures on nuclei.
It should be emphasized that the changes in the electron fraction 
through neutrino interactions can be ignored in the ejecta,
as we explain in \S \ref{sec:neucap}.

In the relatively cool regime of $T < 9 \times 10^9 \K$, 
NSE breaks down and the chemical composition 
is calculated with a large nuclear reaction network~\citep[network B in][]{nishimura06}.
The network includes about 4000 nuclei
from neutron and proton up to fermium, whose atomic number 
$Z = 100$~\citep[see Table 1 of][]{nishimura06}.
The network contains reactions such as 
two and three body reactions, various decay channels, and 
electron-positron capture~\citep[for details, see][]{fujimoto07}.
Experimentally determined masses and reaction rates are adopted if available.
Otherwise, 
theoretical nuclear data, such as nuclear masses, rates of two body reactions,
and $\beta$-decays, 
are calculated with a mass formula based on 
the extended Thomas-Fermi plus Strutinsky integral (ETFSI)~\citep{goriely01}.
Spontaneous and $\beta$-delayed fission processes are taken into accounts
in the network.
We note that the dependence of abundances of \rr-elements on the mass formula
are discussed in \citet{fujimoto07} for model R12.

%%%%%%%%%%%%%%% subsection 3.3
\subsection{Composition of Collapsar Jets} \label{sec:composition}

Once we obtain the density and temperature evolution of a jet particle, 
we can follow its abundance evolution
during infall and ejection, through a post-processing calculation 
using the NSE code and the nuclear reaction network described in the previous subsection. 
We integrate the abundances of all jet particles weighted by their masses to obtain 
the composition of the jets from the collapsars.

The maximum densities and temperatures of the jet particles $\rho_{\rm max}$ and $T_{\rm max}$
(Figure \ref{fig:peak}), are good indicators of the nuclear yields of the ejecta, 
which are experienced explosive nuclear burning~\citep[see Fig. 8 of ][]{thielemann98}.
For $T_{\rm max} < 5 \times 10^9 \K$, 
the particles are abundant in $\alpha$-elements, 
such as \nuc{Si}{28}, \nuc{S}{32}, \nuc{Ar}{36}, and \nuc{Ca}{40},
due to incomplete Si burning or explosive O burning.
On the other hand, for $T_{\rm max} > 5 \times 10^9 \K$
the particles undergo 
complete Si burning, which leads to the synthesis 
of a large fraction of \nuc{Ni}{56}.
The jet compositions for models R10 and S10 are therefore similar to those of ejecta 
from the Fe- and Si-rich layers of Type II SNe.
It should be emphasized that 
the ejected masses for $\alpha$-elements lighter than Fe are
underestimated because ejecta from an outer O-rich layer ($>$ 10,000 \km) are not taken into account.

%%%%%%%%%%%%%%%%%%%%%%%%%%%%%%%%%%%%%%%%%%%%%%%%%%%
%% Figure: maximum densities and Ye
%%%%%%%%%%%%%%%%%%%%%%%%%%%%%%%%%%%%%%%%%%%%%%%%%%%
\begin{figure*}[ht]
 \plottwo{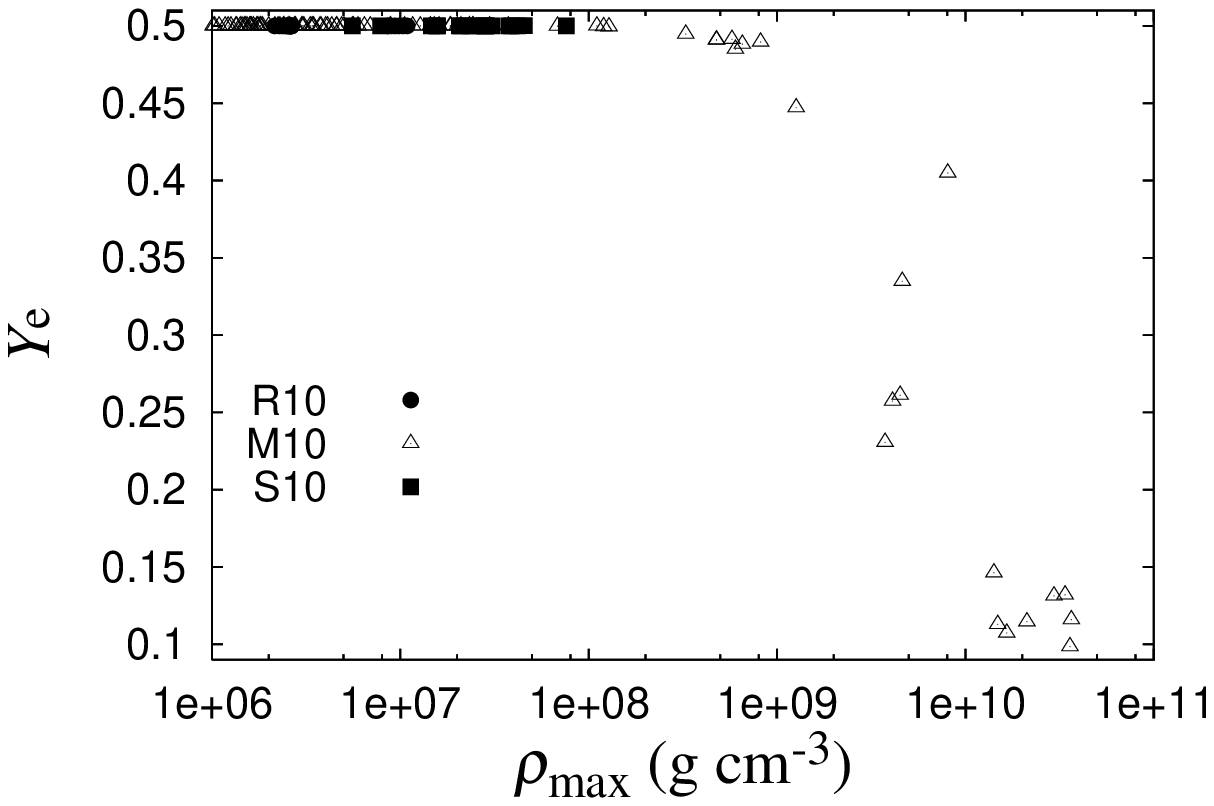}{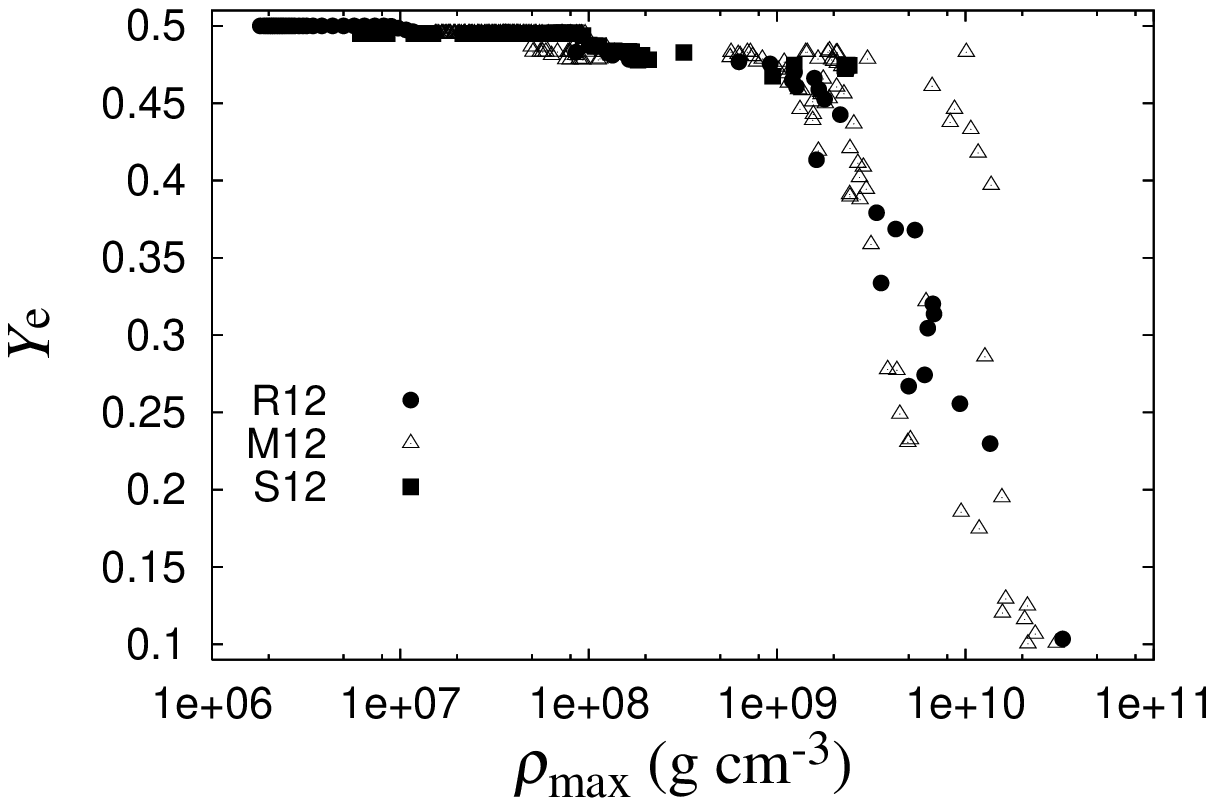}
\caption{
Electron fraction, $Y_e$, at $9 \times 10^{9} \K$
vs. maximum density of jet particles, $\rho_{\rm max}$,
for models with $B_0 = 10^{10}\, \rm G$ (left)
and $B_0 = 10^{12}\, \rm G$  (right).
Circles, triangles, and squares indicate
jet particles for models with rapidly, moderately, and slowly rotating cores, respectively.
} \label{fig:peak-ye}
\end{figure*} 

%%%%%%%%%%%%%%%%%%%%%%%%%%%%%%%%%%%%%%%%%%%%%%%%%%%
%% Figure: Time evolution of Ye,s,tdyn \& T
%%%%%%%%%%%%%%%%%%%%%%%%%%%%%%%%%%%%%%%%%%%%%%%%%%%
\begin{figure*}[ht]
\epsscale{0.8}
\plotone{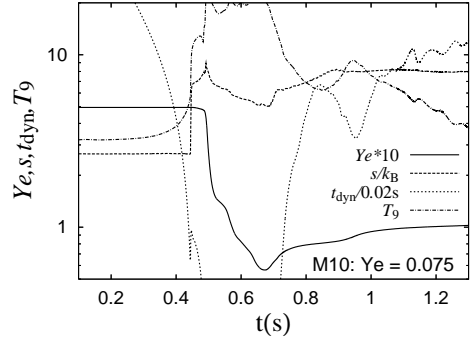}
\caption{
Time evolution of the electron fraction, entropy per baryon, dynamical timescale ($r/v_r$), 
and temperature for the most neutron-rich jet particle in model M10.
} \label{fig:time_evol}
\epsscale{1.0}
\end{figure*} 

%%%%%%%%%%%%%%%%%%%%%%%%%%%%%%%%%%%%%%%%%%%%%%%%%%%
%% Figure: s,tdyn \& Ye at T9 = 9
%%%%%%%%%%%%%%%%%%%%%%%%%%%%%%%%%%%%%%%%%%%%%%%%%%%
\begin{figure*}[ht]
%% \epsscale{0.8}
\plotone{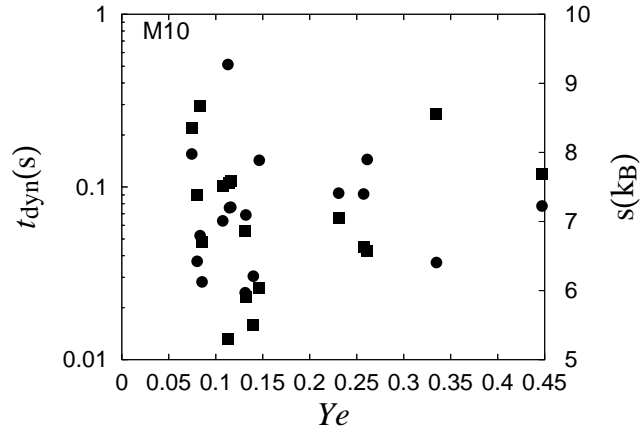}
\caption{
Dynamical timescale (squares) and entropy per baryon (circles) vs. $Y_e$ 
for the neutron-rich jet particles in model M10 at $9 \times 10^9 \rm \,K$.
} \label{fig:s_tdyn-ye}
\epsscale{1.0}
\end{figure*} 

%%%%%%%%%%%%%%%%%%%%%%%%%%%%%%%%%%%%%%%%%%%%%%%%%%
%%% Figure: M(Ye) - Ye
%%%%%%%%%%%%%%%%%%%%%%%%%%%%%%%%%%%%%%%%%%%%%%%%%%%
\begin{figure*}[ht]
\epsscale{1.8} %% for emulate apj on Windows ? 
%%\plottwo{fig3a.eps}{fig3b.eps}
\plottwo{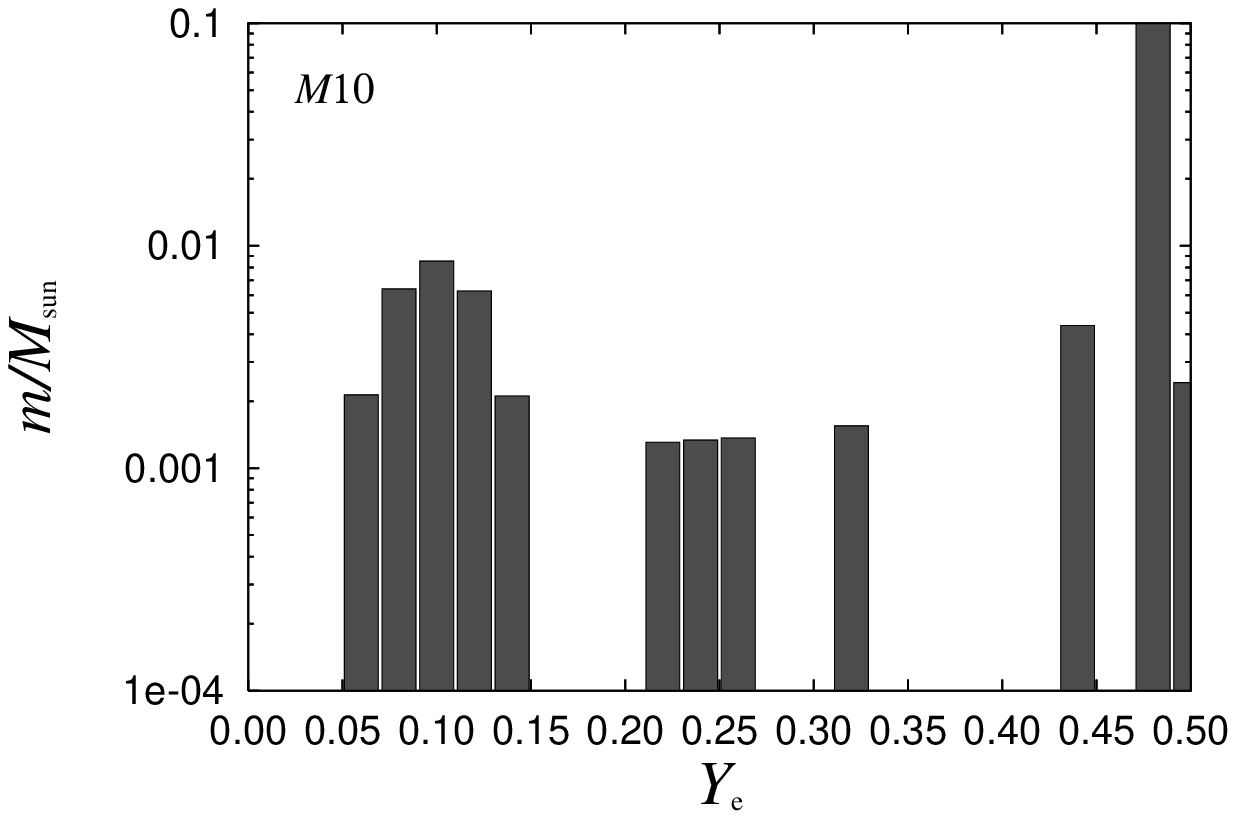}{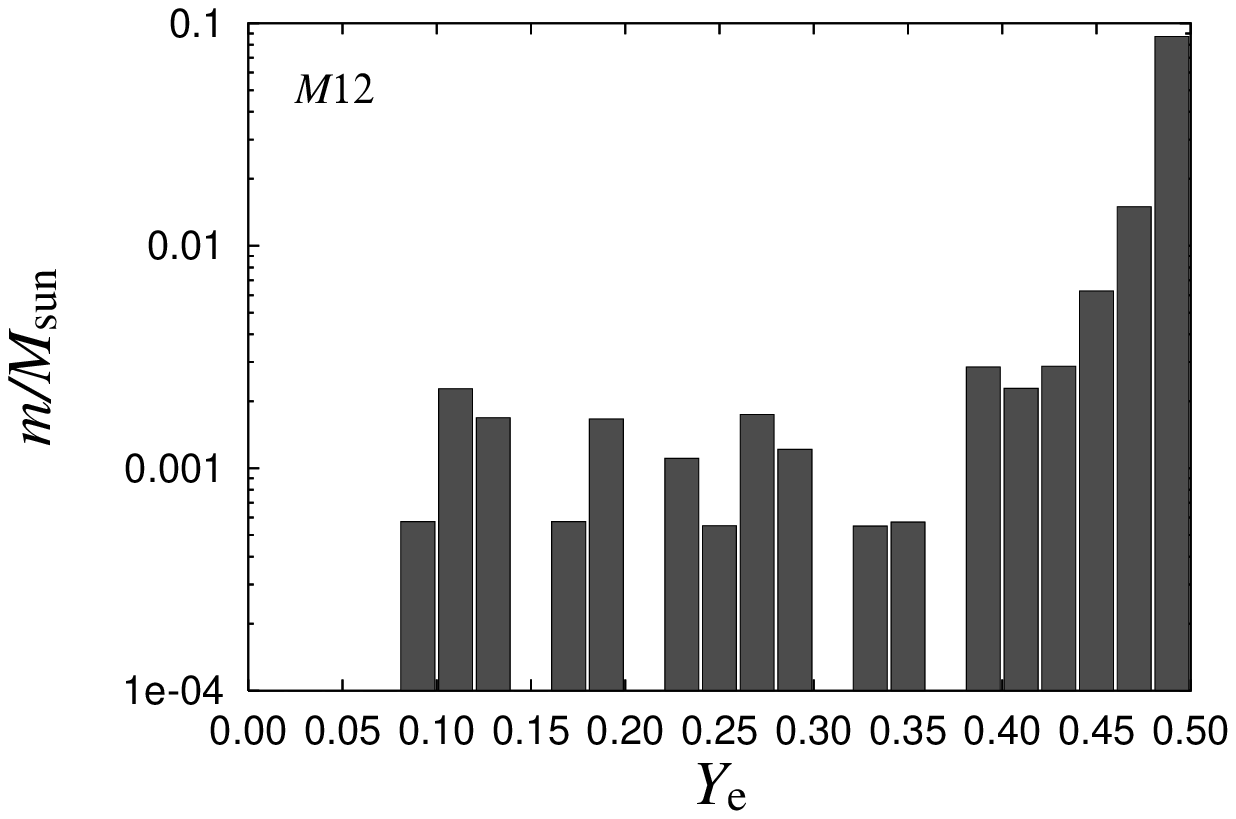}
\caption{
Mass distribution as a function of $Y_e$ in the ejecta for models
M10 (left) and M12 (right).
} \label{fig:mass-ye}
\end{figure*} 

Figure \ref{fig:peak-ye} shows
$Y_e$ with respect to $\rho_{\rm max}$ for the jet particles.
Here $Y_e$ is the electron fraction at $9 \times 10^9\K$ of the the particles.
We note that all ejecta have $Y_e \sim 0.5$ for R10 and S10 
and $Y_e \ge 0.467$ for S12,
while there exist many particles with lower values, $Y_e \le 0.4$,
for M10 and M12, as well as R12.
The \rr-process operates in these low-$Y_e$ or neutron-rich, jet particles.
Figure \ref{fig:time_evol}
shows the time evolution of the electron fraction, entropy per baryon, dynamical timescale, 
which is evaluated as $r/v_r$, 
and temperature for the most neutron-rich jet particle in M10.
Here $v_r$ is the radial velocity of the particle.
As the particle falls in toward the black hole, it passes through a shock surface ($\sim 0.44\s$),
and the temperature and entropy are enhanced as a result of shock heating.
Despite the short dynamical timescale ($< 10^{-4} \rm s$) 
during the infall near the black hole,
the electron fraction decreases through rapid electron capture on protons due to 
the high density and temperature.
This rapid capture leads to fast neutrino cooling, which lowers the entropy 
efficiently ($\sim 0.48-0.68\s$).
After the ejection of the particle through the jet, which may enhance the entropy 
of the particle ($\sim 0.7-0.9\s$), 
the temperature drops to less than $5 \times 10^9$ K
and neutron capture on nuclei proceeds efficiently  
because of the relatively slow ejection ($t_{\rm dyn} > 0.1 \, \rm s$).
The entropy of particles with low $Y_e$ is low  
[$\sim (5-10) k_{\rm B}$, where $k_{\rm B}$ is the Boltzmann constant, 
and the dynamical timescale is relatively long ($\sim 0.01-0.3$ s), 
as shown in Figure \ref{fig:s_tdyn-ye}, 
in which the entropy per baryon and dynamical timescale are shown for $T = 9 \times 10^9 \rm \, K$.
The neutron-rich particles ($Y_e < 0.2$) are likely to have 
a high neutron-to-seed ratio ($> 100$) and a low number of fission cycles, 
as discussed by \citet[][see their Figs. 5 and 6]{meyer97}, 
although they considered cases with much higher entropy [$\sim (50-500) k_{\rm B}$].

Figure \ref{fig:mass-ye} shows
the mass distribution with respect to $Y_e$ in the ejecta for M10 and M12
\citep[see Fig. 9 of][for R12]{fujimoto07}.
We find that the masses of particles with lower $Y_e$ ($\le 0.4$)
are 0.031 and 0.015$\Ms$ for M10 and M12, 
respectively, which are comparable to the value for R12 (0.018$\Ms$).
Note that the ejected masses, $M_{\rm ej}$, are 0.222 and 0.129$\Ms$
through the jets for M10 and M12, respectively (Table 1).
Therefore, large amounts of \rr-elements are ejected through the collapsar jets
for M10 and M12, as in model R12~\citep{fujimoto07}.

\begin{figure*}[ht]
\epsscale{1.8} %% for emulate apj on Windows ? 
\plottwo{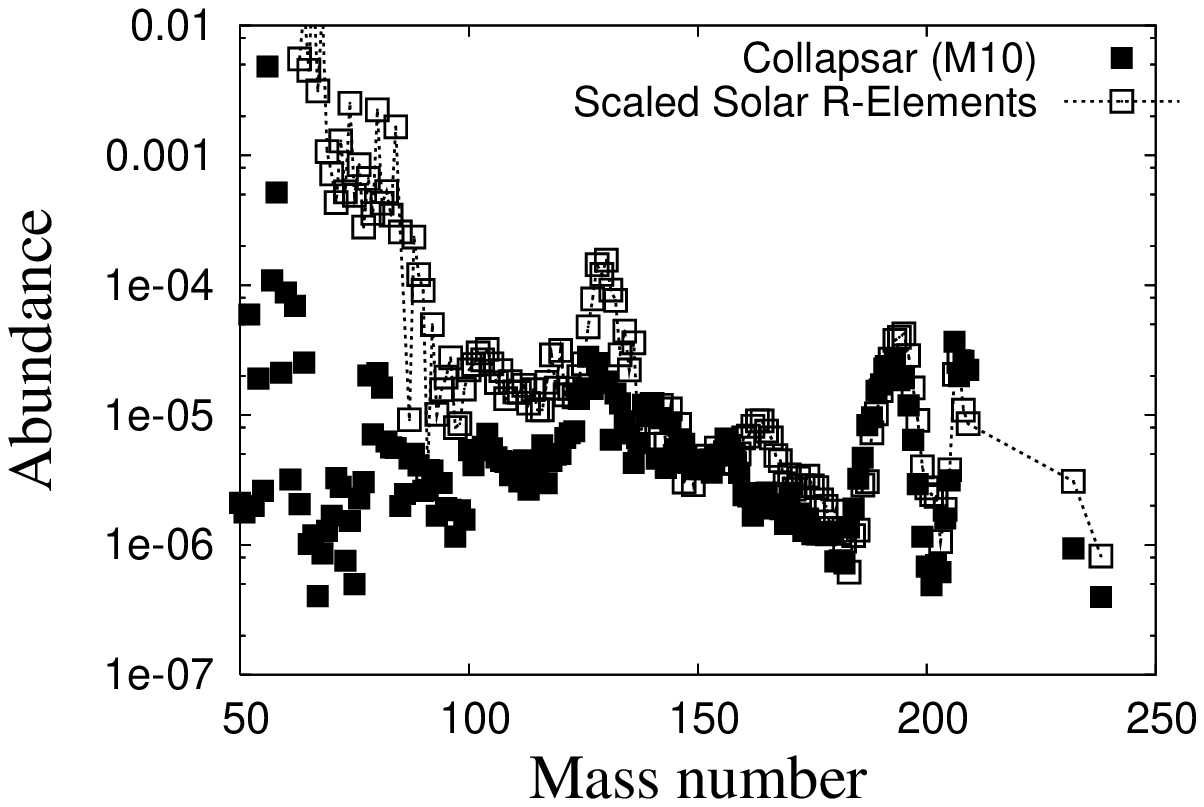}{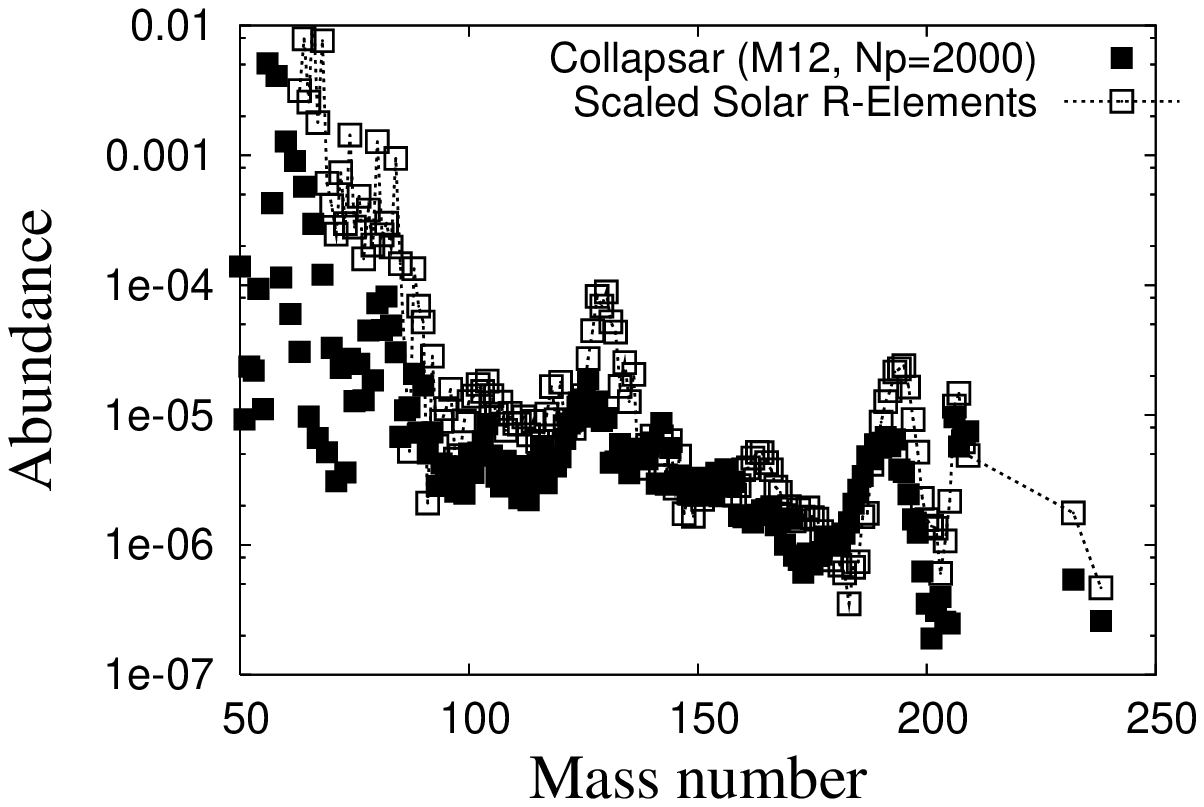}
\caption{
Abundances of jets vs. mass number $A$ for models
M10 (left) and M12 (right).
Filled squares and open squares with dotted line represent 
the collapsar jets and \nuc{Eu}{153}-scaled abundances of 
solar \rr-elements~\citep{arlandini99}, respectively.
} \label{fig:x-a}
\end{figure*}

The compositions of jets 
as a function of mass number, $A$, are shown in Figure \ref{fig:x-a}
for M10 (left) and M12 (right).
The jets from the collapsars of M10 and M12 have abundance profiles 
similar to that of solar \rr-elements, just as for R12~\citep{fujimoto07},
although there are details in the profiles that are rather different from 
that of the solar \rr-elements.
We note that appreciable amounts of U and Th are produced in the jets 
for M10 and M12, as in R12~\citep{fujimoto07}.
We also note that the collapsar jets have abundant \rr-elements compared with Fe,
as we shall see clearly in Figure \ref{fig:M-E} below.
In the collapsar jets, [Eu/Fe] and [Ba/Fe] are $\sim 4$ and $\sim 3$, respectively, 
much larger than the values for the metal poor star, 
HE2148-1247, 
which shows large enhancements of \rr-elements ([Eu/Fe] = 2.0 and [Ba/Fe] = 2.4)~\citep{cohen03}.
Here [A/B] is $\log(Y_{\rm A}/Y_{\rm B})- \log(Y_{\rm A}/Y_{\rm B})_{\odot}$ for elements A and B with number fractions $Y_A$ and $Y_B$.

%%%%%%%%%%%%%%%%%%%%%%%%%%%%%%%%%%%%%%%%%%%%%%%%%%
%%% X-A
%%%%%%%%%%%%%%%%%%%%%%%%%%%%%%%%%%%%%%%%%%%%%%%%%%
%%% Figure: Integrated abundances ETFSI

%%%%%%%%%%%%%%%%%%%%%%%%%%%%%%%%%%%%%%%%%%%%%%%%%%%%%%%%%%%%%%%
%%%% P-nuc. in M10 \& M12
For M10 and M12, the collapsar jets have large amounts of light \pp-nuclei
(\nuc{Se}{74}, \nuc{Kr}{78}, \nuc{Sr}{84}, and \nuc{Mo}{92})
and heavy {\it p}-nuclei (\nuc{In}{113}, \nuc{Sn}{115}, and \nuc{La}{138}).
These \pp-nuclei are much more abundant than is found in core-collapse SNe and
are produced without {\it s}-process seeds, as in model R12~\citep{fujimoto07}.

\begin{figure*}[ht]
\epsscale{0.8}
%%\plotone{b12omg2.5-lres-np1e3-dist.eps}
\plotone{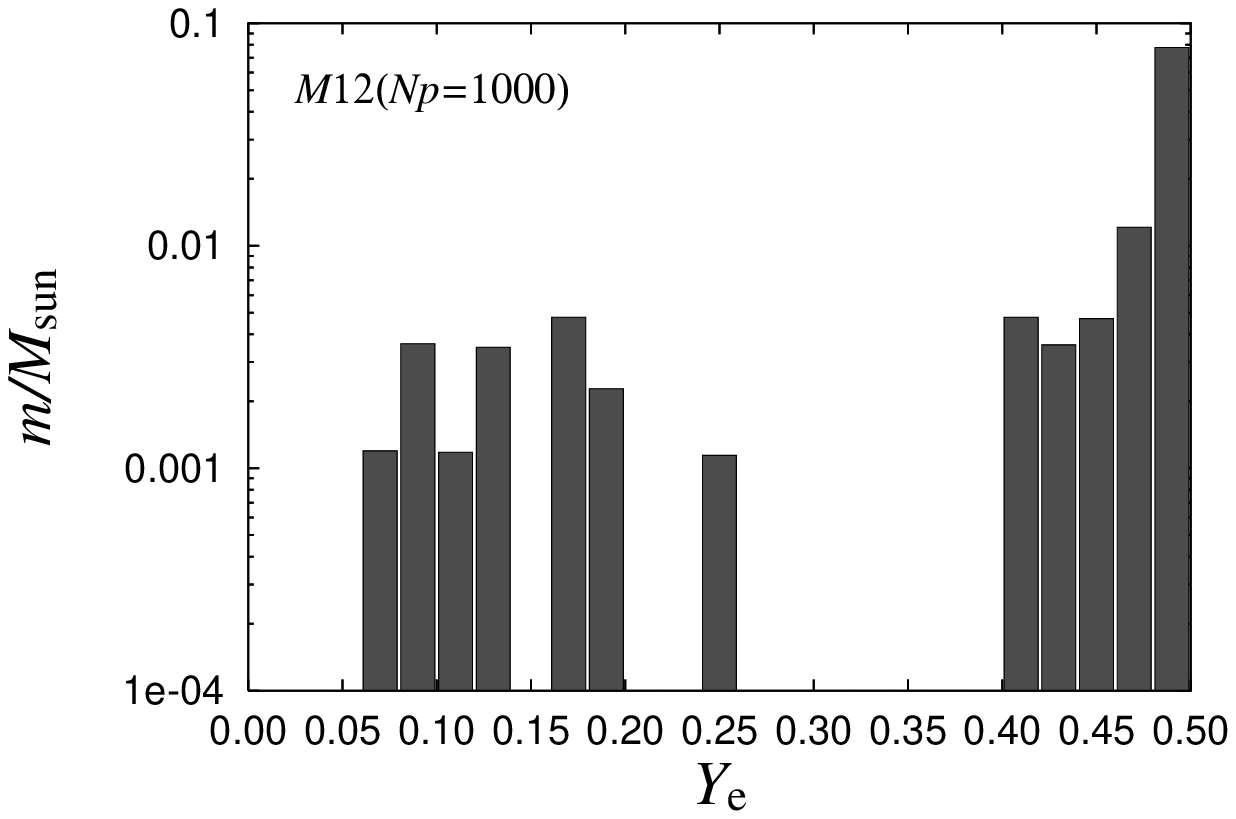}
\caption{
Same as Figure \ref{fig:mass-ye} but for model M12 
with half the number of particles ($N_p = 1000$).
} \label{fig:mass-ye-m12np1000}
\epsscale{1.0}
\end{figure*}

%%%%%%%%%%%%%%%%%%%%%%%%%%%%%%%%%%%%%%%%%%%%%%%%%%%%%%%%%%%%%%%
%%%% Np, mashes dependences
The ejected mass slightly increases for larger $N_p$.
The values of $M_{\rm ej}$ are 0.222, 0.121, and 0.084 $\Ms$
for M10, M12, and R12, respectively, with $N_p = 1000$.
When we use 2000 tracer particles,
these increase to 0.245, 0.129, and 0.124 $\Ms$
for M10, M12, and R12, respectively.
The mass distribution with respect to $Y_e$ also depends weakly on $N_p$ for $Y_e < 0.4$,
and thus the abundance profiles of \rr-elements change slightly for different values of $N_p$.
Figure \ref{fig:mass-ye-m12np1000} shows 
the mass distribution with respect to $Y_e$ in the ejecta of M12 
with half the number of particles ($N_p = 1000$).
Compared with $N_p = 2000$ case (Figure \ref{fig:mass-ye}, right), 
ejecta with $Y_e \sim 0.3$, in which nuclei are synthesized mainly with $A$ around 130
(see Figure 4 of \citep{fujimoto07}) are deficient.
The \rr-elements with $A \sim 130$ are less abundant than 
for the case with $N_p = 2000$ (Figure \ref{fig:x-a}, right).
We conclude that the abundance profiles of the jets do not depend strongly on $N_p$.
We note that the profiles also depend weakly on 
the nuclear mass formula and the numerical resolution of the MHD simulations,
as discussed in \citet{fujimoto07}.

%%%%%%%%%%%%%%%%%%%%%%%%%%%%%%%%%%%%%%%%%%%%%%%%%%
%%% M(Ye) - Ye: M12 for Np = 1000

\subsection{Masses and Energies of Collapsar Jets}

\begin{figure*}[ht]
%% \epsscale{0.8}
%\epsscale{0.5} %% for draft
%%\plotone{fig5.eps}
\plotone{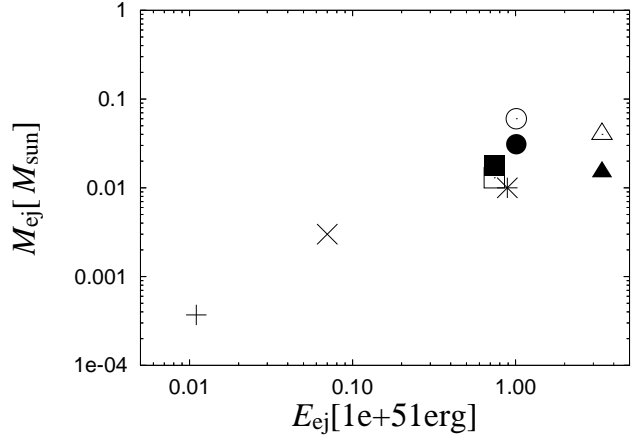}
\caption{
Masses of \nuc{Ni}{56} and \rr-elements vs. energy in the jets.
The plus sign, open square, open circle, open triangle, cross, and asterisk indicate
\nuc{Ni}{56} for models R10, R12, M10, M12, S10, and S12, respectively.
The filled square, filled circle, and filled triangle show
\rr-elements for R12, M10, and M12, respectively.
We set the masses of \rr-elements to be those of particles with $Y_e \le 0.4$.
Particles with $Y_e \le 0.4$ cannot be ejected for models R10, S10, and S12.
} \label{fig:M-E}
\end{figure*}

%%%%%%%%%%%%%%%%%%%%%%%%%%%%%%%%%%%%%%%%%%%%%%%%%%
%%% Mej vs Eej
%%%%%%%%%%%%%%%%%%%%%%%%%%%%%%%%%%%%%%%%%%%%%%%%%%
The jets are abundant in \nuc{Ni}{56} because of the complete Si burning, 
as described in \S \ref{sec:composition} (Figure \ref{fig:peak}).
Figure \ref{fig:M-E} shows the mass of \nuc{Ni}{56}
with respect to the energy of the jets, 
where the energies are the sum of the kinetic energy $(E_k)_{\rm ej}$
and the internal energy $(E_i)_{\rm ej}$ of all the jet particles.
The plus sign, open square, open circle, open triangle, cross, and asterisk indicate
\nuc{Ni}{56} for models R10, R12, M10, M12, S10, and S12, respectively.
We find that the \nuc{Ni}{56} masses are roughly proportional to the energies of the jets.
High energy jets (M10, R12, M12, and S12) could be observed as
GRBs with normal SNe if the collapsars are located near the Earth.
On the other hand, 
low energy jets (R10 and S10) might be observed as GRBs without SNe, 
like the recently observed nearby GRBs, 060505 and 060614~\citep{fynbo06,gehrels06}, 
even if the jets are accompanied by GRBs located near Earth.

The ejected masses of \nuc{Ni}{56} in our simulations are lower than
those in the hydrodynamic simulations of \citet{mw99}, in which 
a large amount of \nuc{Ni}{56} is expected to be ejected through winds driven by viscous heating.
In contrast, in our MHD simulations, 
\nuc{Ni}{56} is ejected through the jets, 
not through winds, which disappear in the simulations.
The viscous heating is implemented with the $\alpha$-prescription
in MacFadyen \& Woosley's hydrodynamic code
while such heating is not taken into account in our MHD simulations.
As a result of viscous heating, the entropy of the winds is greater than $(20-30) k_{\rm B}$ 
per baryon and is much higher than 
that of the jets in our MHD simulation (Fig. \ref{fig:time_evol}).
We note that 
winds have not appeared in MHD simulations by other groups~\citep{proga03,nagataki07} 
or in the hydrodynamic simulation with a low $\alpha \sim 0.001$ by \citet{mw99}.
The winds are therefore generated by an ``$\alpha$-viscosity'' and may be artificial.

%%%%%%%%%%%%%%%%%%%%%%%%%%%%%%%%%%%%%%%%%%%%%%%%%%
%%%% GRB jets
%%%%%%%%%%%%%%%%%%%%%%%%%%%%%%%%%%%%%%%%%%%%%%%%%%
It should be emphasized that the velocities of the jets are less than 
$0.2 c$, where $c$ is the speed of light, in our MHD simulations.
The jets are therefore too slow to produce GRBs.
We note, however, that the low-density polar region near the black hole in the collapsars
is a possible candidate for the production site of GRB jets.
This is because the region is located near the inner region of the accretion disk,
in which a large magnetic energy $>10^{51} \ergs$ is sustained
and through which mass accretion takes place at rates above $0.05 \Ms\psec$~\citep{fujimoto06}.
The energy liberated in the disk can be greater than $9 \times 10^{51} \ergs\psec$
even for the low efficiency ($= 0.1$) of energy liberation.
In addition, the energy of neutrinos, 
which emanates from the disk at rates higher than $5 \times 10^{51} \ergs\psec$~\citep{fujimoto06},
is partly transfered to the polar region from the disk,
and acts to heat the polar region~\citep{nagataki07}.

%%%%%%%%%%%%%%%%%%%%%%%%%%%%%%%%%%%%%%%%%%%%%%%%%%%%%%%%%%%%%%%%%%%%%%%%%%%
%%%% Soderberg two components jets : GRBs and Ni56
%%%%%%%%%%%%%%%%%%%%%%%%%%%%%%%%%%%%%%%%
Moreover,
collapsar jets may have two components: ultra relativistic, baryon-poor jets and 
non relativistic, baryon-rich jets~\citep{soderberg06}.
SNe associated with GRBs are perhaps induced through an aspherical, jet-like explosion,
which produces non relativistic, baryon-rich jets~\citep{maeda03}.
The SNe accompanying GRBs suggest that the jets have two components. 
The jets in the present study correspond to the non-relativistic, baryon-rich jets.

The masses of \rr-elements are also revealed in Figure \ref{fig:M-E}, 
with the filled square, filled circle, and filled triangle standing for models
R12, M10, and M12, respectively.
The \rr-process is found to be operate only in energetic jets ($> 10^{51}$ \ergs), 
which have considerable amounts of \nuc{Ni}{56}.
The energetic explosion leads to the ejection of not only 
a large amount of \nuc{Ni}{56} but also considerable amounts of 
neutron-rich material, on which the \rr-process operates, near the black hole.
If jets are realized with higher energies in a collapsar, 
they probably have larger amounts of \nuc{Ni}{56} 
as well as appreciable amounts of \rr-elements.
The jets can be observed as GRB associated with a hypernova, 
in which abundant \rr-elements are possibly produced.
Therefore, GRBs with hypernovae, such as GRB 980425 and GRB 030329, could eject 
appreciable amounts of \rr-elements in addition to a large amount of \nuc{Ni}{56},
if the hypernova is induced by jets that are driven near the black hole.

%%%%%%%%%%%%%%%%%%%%%%%%%%%%%%%%%%%%%%%%
%%% Figure: Mej vs Eej
%%%%%%%%%%%%%%%%%%%%%%%%%%%%%%%%%%%%%%%%

%%%%%%%%%%%%%%%%%%%%%%%%%%%%%%%%%%%%%%%%%%%%%%%%%%
%%% Figure: Time evolution of Mdot_ej, \& Mej 
%%%%%%%%%%%%%%%%%%%%%%%%%%%%%%%%%%%%%%%%%%%%%%%%%%
\begin{figure*}[ht]
%% \epsscale{0.8}
%\epsscale{1.8} %% for emulate apj on Windows ? 
\plotone{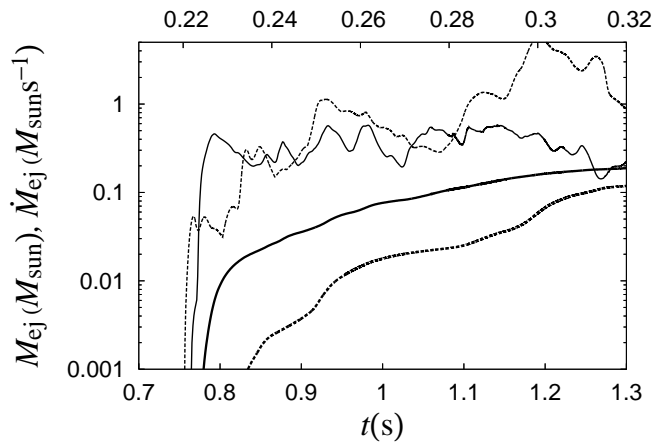}
\caption{
Time evolution of the mass ejection rate $\dot{M}_{\rm ej}$ (thick line) and integrated mass of ejecta $M_{\rm ej}$  (thin line)
after launching of the jets for models M10 (0.7-1.3 s; solid lines) and M12 (0.21-0.32 s; dashed lines). 
} \label{fig:mej-dmej}
\end{figure*} 

We terminated the simulations at $t = t_f$ (Table 1), which is much shorter than the typical duration $\sim 10\s$
of a long GRB.
If we perform the simulations for a much longer time, the ejected masses and energies will increase.
Figure \ref{fig:mej-dmej} shows 
time evolution of mass ejection rates, $\dot{M}_{\rm ej}$, and integrated mass, $M_{\rm ej}$, 
ejected through a spherical surface with a radius of 1000\km, 
after the launch of the jets for models M10 (0.7-1.3 $\s$) and M12 (0.21-0.32 $\s$).
Mass ejection takes place at rates of 0.2-0.5 $\Ms\,\rm s^{-1}$ for M10,
while the jets for M12 have much higher ejection rates, up to $5 \Ms\,\rm s^{-1}$.
The average ejection rates from $t_{\rm jet}$ to $t_{f}$ (Table 1) are 0.42 and 1.17 $\Ms\,\rm s^{-1}$ 
for M10 and M12, respectively.
If these high ejection phases were to continue for 10$\s$, the ejected masses could be
4.2 and 11.7 $\Ms$ for M10 and M12, respectively.
It should be emphasized that the GRB duration indicates the continuous ejection of
ultra-relativistic, baryon-poor jets, not baryon-rich ones.
The mass fraction of \nuc{Ni}{56} in the ejecta is about 0.3 for M10 and M12.
The masses of \nuc{Ni}{56}, therefore, can be comparable to those of hypernovae 
for the ejection of 1-2$\Ms$ material through the collapsar jets, 
which is realized in the ejection of the jets during a period of 2.5-5 $\s$ and 1-2 $\s$
for the high ejection phases in M10 and M12, respectively.
These \nuc{Ni}{56}-rich ejecta could have 
large amounts of \rr-elements, $\sim 0.1-0.2 \Ms$, 
because of the large abundances ($\sim 0.1$ in mass) of the elements for M10 and M12.
We need, however, much longer term simulations to determine the duration of 
the ejection of baryon-rich jets and the masses of \nuc{Ni}{56} and \rr-elements.

%%%%%%%%%%%%%%%% Section 
\section{Discussion} \label{sec:discussion}

\subsection{Effects of Neutrino Captures}\label{sec:neucap}

We first discuss the effects of neutrino capture on the composition of the collapsar jets.
As shown in \citet{fujimoto07}, jet particles with lower $Y_e$ have more abundant 
heavy nuclei.
If the neutrino capture on nucleons is taken into account, 
the electron fraction of the particles changes compared with that in the present study, 
so that heavy nuclei may be less abundant.
Taking into account the change in electron fraction due to neutrino captures on nucleons,
we follow the evolution of the electron fraction of the jet particles for model M10,
in which \rr-elements are abundantly ejected through the jets 
(Figure \ref{fig:x-a}, left).
Then we estimate the difference in $Y_e$ between cases with and without neutrino capture
for all the jet particles.
We assume that neutrino emission is isotropic for simplicity,
although it actually is aspherical because neutrinos are 
chiefly emitted from the dense and hot disk~\citep[Fig.9 of][]{fujimoto06}.
We evaluate the capture rates of electron neutrinos and anti-neutrinos, 
$\lambda_{\nu_e}$ and $\lambda_{\bar{\nu}_e}$, as
\begin{eqnarray}
\lambda_{\nu_e} &=& 4.83 Y_n L_{\nu_e,\,51} E_{\nu_e}/r_6^2, \\
\lambda_{\bar{\nu}_e} &=& 4.83 Y_p L_{\bar{\nu_e},\,51} E_{\bar{\nu_e}}/r_6^2,
\end{eqnarray}
where $Y_n$, $Y_p$, $L_{\nu_e,\,51}$, $L_{\bar{\nu}_e,\,51}$,
$E_{\nu_e}$, $E_{\bar{\nu_e}}$, and $r_6$ are respectively
the number fractions of neutrons and protons,
the luminosities of electron neutrinos and anti-neutrinos in units of $10^{51}\erg\psec$, 
the average energies of electron neutrinos and anti-neutrinos in $\mev$, and
the distance between a particle and the center of the collapsar 
in units of $10^6\cm$~\citep{qw96}.
We set $E_{\nu_e}$ and $E_{\bar{\nu_e}}$ to 10 and 15, respectively.
We have evaluated the neutrino luminosity with the two-stream approximation~\citep{dpn02} in our MHD simulations.
However, 
the luminosity is the sum of neutrino luminosities over all neutrino flavors, 
and thus $L_{\nu_e,\,51}$ and $L_{\bar{\nu}_e,\,51}$ are not provided 
by the two-stream approximation.
We therefore estimate $L_{\nu_e,\,51}$ and $L_{\bar{\nu}_e,\,51}$ 
according to Appendix B of \citet{ruffert96},
with the density and temperature profiles obtained from
the MHD simulations.
We note that the evaluated total luminosities are comparable
with the above two methods.
We find that $L_{\nu_e,\,51}$ and $L_{\bar{\nu}_e,\,51}$ 
are less than 3.6 and 3.1, respectively, for M10, 
and that $Y_e$ increases compared with the case without the neutrino capture,
especially for particles with $Y_e < 0.2$.
However, the differences in $Y_e$ are small, up to 0.03.
There exist particles with $Y_e < 0.15$, and 
the \rr-process operates in the jets to produce U and Th abundantly.
We note that for some low $Y_e$ particles, 
$Y_e$ decreases compared with the case without neutrino capture.

Moreover, we followed the evolution of the electron fractions of jet particles for model M10,
with $L_{\nu_e,\,51}$ and $L_{\bar{\nu_e},\,51}$ set 5 times larger 
than the luminosities evaluated with the results of the MHD simulation.
This is because when the inner boundary 
of the computational domain for the MHD simulations, $r_{\rm in}$, is set to $10\km$
instead of $50\km$ as in the present study, 
the neutrino luminosity known to increase slightly~\citep{fujimoto06}.
Aspherical neutrino emission from the disk may increase 
the neutrino intensity on the jet particles effectively.
We find that the difference in $Y_e$ is as much as 0.12
and that there are several particles with low $Y_e < 0.2$, 
in which U and Th are abundantly synthesized.
The total mass of the particles amounts to 0.011$\Ms$.
We conclude that
the \rr-process probably operates in the collapsar jets to synthesize considerable amounts of U and Th
even if we take into account neutrino capture on nucleons and aspherical neutrino emission.

In addition to the changes in electron fraction due to neutrino capture, 
neutrino interactions are important for the dynamics of collapsar jets 
through neutrino heating~\citep{nagataki07}.
The heating from neutrino captures and annihilations 
between neutrinos and anti-neutrinos may increase the energy of the jets.
The amount of \nuc{Ni}{56} ejected through the jets is likely to increase 
and could produce hypernovae with appreciable amounts of \rr-elements.

\subsection{Ejecta from the Outer Layers of Collapsars} \label{sec:outer}

In the present study, 
we have calculated the masses and composition of the jets 
ejected from the inner region ($\le$ 10,000\km) of the collapsars.
The ejecta from the outer layers ($>$ 10,000\km) have a large amount of material and 
are abundant in elements lighter than Fe.

%%%%%%%%%%%%%%%%%%%%%%%%%%%%%%%%%%%%%%%%%%%%%%%%%%%%%%%%%%%%%%%%%%%%%%%%%%%
It should be emphasized that comparable amounts of \nuc{Ni}{56} may be
ejected through the jets from outer O-rich layers ($>$ 10,000 $\km$), 
which are not covered by the computational domain of the present MHD simulations.
We have evaluated $M($\nuc{Ni}{56}$)_{\rm out}$ 
for ejecta from the core-collapse SN of a 40 $\Ms$ star, 
which is the initial model for our MHD simulations, 
using the spherical explosion model~\citep{hashimoto95}.
Here $M($\nuc{Ni}{56}$)_{\rm out}$ 
is the mass of \nuc{Ni}{56} ejected from the outer layers ($>$ 10,000\km).
We set the explosion energy of the SN to be $1.0 \times 10^{51} \ergs\psec$, 
which is comparable to the energy of the ejecta from the collapsar of model M10.
We find that $0.115\Ms$ of \nuc{Ni}{56}is ejected from the outer layers.
If we take into account an aspherical jet-like explosion, 
$M($\nuc{Ni}{56}$)_{\rm out}$ is estimated to be $\sim 0.04 \Ms$.
This is because, for M10, the jets have a relatively large opening angle of $\sim 30$ degree.
We note that 
$M($\nuc{Ni}{56}$)_{\rm out}$ is comparable to the mass of \nuc{Ni}{56}
from the inner layers $\le$ 10,000\km.

As discussed in \citet{fujimoto07}, 
the \pp-process in O-rich layers is expected to operate in collapsars, 
as in the \pp-process layers (PPLs) of core-collapse SNe~\citep{rayet95}.
This is because {\it s}-process seeds for the \pp-process exist in the layers, 
as a result of the weak {\it s}-process operating during He core burning of the stars, 
and the jets propagate in the layers to heat them to high enough temperatures for the \pp-process to operate efficiently. 
Therefore, the masses of \pp-nuclei can be enhanced by means of the \pp-process. 
The \pp-nuclei ejected from collapsars with high energy jets represent the sum of those in the ejecta from oxygen-rich layers, 
with composition similar to that in the PPLs, and those in the jets from the inner core, which abundantly contain \pp-nuclei that are deficient in the SNe,
such as \nuc{Mo}{92}, \nuc{In}{113}, \nuc{Sn}{115}, and \nuc{La}{138}.

\subsection{Effects of Nuclear Energy Generation}

\begin{figure*}[ht]
%% \epsscale{0.8}
%\epsscale{0.5} %% for draft
\plotone{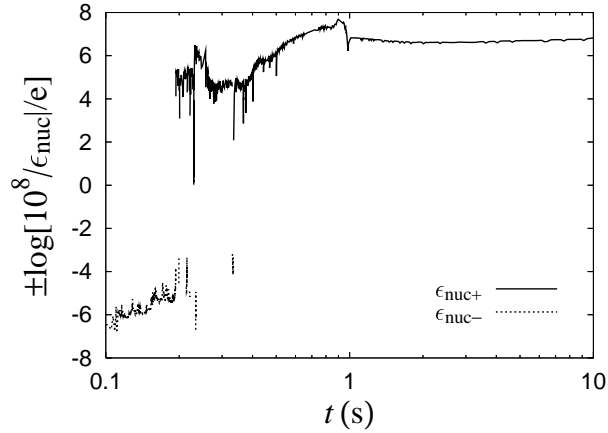}
\caption{
Time evolution of the ratio of the specific energy liberated through nuclear reactions, 
$\epsilon_{\rm nuc}$, to the specific energy, $e$, of the particle 
with $Y_e \sim 0.1$ for model R12.
The logarithm of the magnitude of the ratio times $10^8$, $\log ( 10^8 |e_{\rm nuc}/e| )$, 
is shown by the solid line ($\epsilon_{\rm nuc} \ge 0$),
and is shown by the dashed line $-\log ( 10^8 |e_{\rm nuc}/e| )$ (otherwise). 
Note that $|e_{\rm nuc}|$ becomes greater than $e$ for $|\log ( 10^8 |e_{\rm nuc}/e| )| > 8$,
so that $|e_{\rm nuc}|$ is always less than $e$.
} \label{fig:enuc}
\end{figure*}

We finally discuss how the energy that is generated through nuclear reactions
affects the dynamics of collapsars.
We evaluate the ratio, $R_{\rm nuc}$, of the specific energy liberated in nuclear reactions, 
$\epsilon_{\rm nuc}$, to the specific energy, $e$, of a particle.
Here $e$ is obtained from the results of the MHD simulations and 
$\epsilon_{\rm nuc}$ is calculated as
\begin{equation}
 \epsilon_{\rm nuc} = 9.65 \times 10^{17} \,\sum_i \,[ \,Q_i \,\delta{Y_i} ] \,\, {\rm \erg \,g^{-1}\psec}, 
\end{equation}
where the sum is performed over all the nuclei in the reaction network and
$Q_i$ and $\delta Y_i$ are the mass excess (in \mev) and 
the change in the number fraction of the $i$ th nuclide, respectively.
We obtain $\delta Y_i$ from the abundance evolution of the particle.
%followed with the NSE code and the nuclear reaction network.
We find that $|R_{\rm nuc}|$ is always less than 0.05 in jet particles,
other than neutron-rich jet particles with $Y_e < 0.4$.
Figure \ref{fig:enuc} shows the time evolution of the magnitude of $R_{\rm nuc}$ for
a jet particle with $Y_e \sim 0.1$ in model R12.
Solid and dotted lines represent cases with positive and negative $R_{\rm nuc}$, respectively.
As the jet particle falls in near the black hole ($t \le 0.2\s$), 
heavy elements in the particle are chiefly destroyed to \nuc{He}{4} and then to protons and neutrons
by photodisintegrations, which are endothermic.
On the other hand, during ejection through the jets ($0.2 \s \le t \le 1\s$),
nucleons in the particles recombine to \nuc{He}{4} and then 
heavier nuclei, chiefly neutron-rich nuclei, by means of the \rr-process.
The nuclear energy liberated by the recombination heats the particle
because of the exothermic nature of the recombinations.
After the freeze-out of the \rr-process due to neutron consumption, 
abundant neutron-rich unstable nuclei begin to decay 
through sequences of $\beta$-decays ($t \ge 0.4 \s$).
The liberated energy is comparable to the particle energy, but
$|R_{\rm nuc}|$ is always less than 0.487.
We conclude that the nuclear energy is unlikely to affect the dynamics of the collapsar jets,
significantly, 
as in supernova ejecta with a neutron-to-proton ratio of 10~\citep[see, Fig.3 of ][]{sato74} and
in ejecta with an electron fraction of 0.12 from neutron star mergers~\citep[see, Fig.2 of ][]{frt99}.

%%%%%%%%%%%%%%%%%%%%%%%%%%%%%%%%%%%%%%%%
%%% Figure: Enuc(t)

%%%%%%%%%%%%%%% Summary
\section{Summary and Conclusions}

We have calculated the composition of magnetically-driven jets
ejected from collapsars, or, rapidly rotating massive (40$\Ms$) stars,
based on long-term, magnetohydrodynamic simulations of the stars during core collapse.
The magnetic fields of the collapsars prior to the collapse
are set to be uniform and parallel to the rotational axis.
We consider three sets of angular velocity distributions, 
with rapidly, moderately, and slowly rotating cores, 
and three cases with magnetic fields of $10^8$, $10^{10}$, and $10^{12}$ G.
We follow the evolution of the abundances of about 4000 nuclei in the jets
from the collapse phase to the ejection phase
through the jet generation phase with the aid of an NSE code and 
a large nuclear reaction network.
We summarize our conclusions as follows:

\begin{enumerate}
 \item The collapsars eject appreciable \nuc{Ni}{56}, in amounts 
       comparable to or somewhat less than that from normal core-collapse SNe.
       The ejected \nuc{Ni}{56} masses, which range from $3.7 \times 10^{-4}$ to 0.06$\Ms$, through the jets are roughly proportional to the jet energy.
 \item Less energetic jets that eject small amounts of \nuc{Ni}{56}
       could induce GRBs without a supernova, such as GRB 060505 and GRB 060614.
 \item The \rr-process operates only in energetic jets ($> 10^{51} \ergs$),
       even if changes in the electron fraction due to neutrino captures are taken into account.
       U and Th are synthesized in jets even for a collapsar with a moderately rotating core of $2.5 \, \rm rad \,s^{-1}$
       and magnetic field of $10^{10}\rm\,G$.
 \item The abundance patterns of the energetic jets
       are similar to those of the solar \rr-elements,
       while low-energy jets have compositions similar to the ejecta 
       from core collapse SNe.
 \item Considerable amounts of \rr-elements could be ejected from GRB with a hypernova, 
       if both GRB and hypernova are induced by jets that are driven near the black hole.
 \item Energy released through nuclear reactions is not significant for the dynamics
       of the collapsar jets.
\end{enumerate}

%%%%%%%%%%%%%%%%%%%%%%%%%%%%%%%%%%%%%%%%%%%%%%%%%%%%%%%%%%%%%%%%%%%%%%%%%%%
%%%% Final remarks
In the present study, 
we have calculated the masses and composition of the jets 
ejected from the inner regions ($\le$ 10,000\km) of the collapsars.
The ejecta from the outer layers ($>$ 10,000\km) contain a large amount of mass and 
are abundant in elements lighter than Fe.
As discussed in \S \ref{sec:outer}, the amount of \nuc{Ni}{56} in the ejecta from the outer layers is
comparable to that from the inner layers, 
and considerable amounts of \pp-nuclei are ejected through the jets from the outer layers.
We need to take into account the ejecta from the outer layers to
estimate the masses and abundances of elements lighter than Fe, as well as \pp-nuclei.
This will be our next undertaking.

Relativistic effects related to black hole spin could induce vary energetic jets 
at a later time when the rotation of the black hole becomes rapid 
as a result of continuous mass accretion with high angular momentum 
through than accretion disk.
Amplification of toroidal magnetic fields by the magneto-rotational instability
might also produce more energetic jets if we perform a three-dimensional MHD simulation, 
rather than two-dimensional simulations as in the present study, in particular, for models 
with low-energy jets, such as R10 and S10.
In addition, the standing accretion shock instability, 
whose importance has been recognized in SN explosion~\citep{blondin03}
might change the energies and masses of the jets.

%%%%%%%%%%%%%%%%%%%%%%%%%%%%%%%% Acknowledgments
\acknowledgments{
We thank the anonymous referees for valuable comments
that were helpful in improving the manuscript.
S. F. is grateful to K. Takahashi for fruitful discussions.
This work was supported in part by a 
Grant-in-Aid for Scientific Research from the Ministry of
Education, Culture, Sports, Science and Technology of Japan (No. 17540267).
}

%%%%%%%%%%%%%%%%%%%%%%%%%%%%%%%% References
%%%%%%%%%%%%%%%%%%%%%%%%%%%%%%%% References

%%%%%%%%%%%%%%%%%%%%%%%%%%%%%%%%%%%%%%%%%% Table 1
%%% 

%%% check
%%% units in Table --> 2nd lines

\begin{deluxetable} {cccccccccc}
\tablewidth{0pc}
\tablecaption{Jet Properties}
\tablehead{
\colhead{model} &
\colhead{$B_0$} & 
\colhead{$\Omega_0$} & 
\colhead{$R_0$} &
\colhead{$t_{f}$} &
\colhead{$t_{\rm jet}$} &
\colhead{$M_{\rm ej}$} & 
\colhead{$(E_m)_{\rm ej}$} &
\colhead{$(E_k)_{\rm ej}$} &
\colhead{$(E_i)_{\rm ej}$}
}
\startdata
 R10 & $10^{10}$ & 10  & 1000 & 2.62 & 2.58 & 0.0010 & 2.89e-4 & 0.0274 & 0.0840 \nl     
%% M10 & $10^{10}$ & 2.5 & 2000 & 1.30 & 0.80 & 0.222  & 0.787   & 2.206  & 7.926 \nl
 M10 & $10^{10}$ & 2.5 & 2000 & 1.30 & 0.77 & 0.222  & 0.787   & 2.206  & 7.926 \nl
 S10 & $10^{10}$ & 0.5 & 5000 & 1.34 & 1.30 & 0.0053 & 0.045   & 0.138  & 0.606 \nl
 R12 & $10^{12}$ & 10  & 1000 & 0.36 & 0.20 & 0.083  & 0.097   & 4.58   & 2.79  \nl
 M12 & $10^{12}$ & 2.5 & 2000 & 0.33 & 0.22 & 0.129  & 0.783   & 24.5   & 9.44  \nl
 S12 & $10^{12}$ & 0.5 & 5000 & 0.28 & 0.25 & 0.033  & 0.014   & 3.57   & 5.27  \nl
% R10 & $10^{10}$ & 10  & 1000 & 2.62 & 2.58 & 0.0018 & 4.83e-6 & 0.0142  & 0.0323 \nl
% S10 & $10^{10}$ & 0.5 & 5000 & 1.34 & 1.30 & 0.0065 & 0.046   & 0.468   & 0.603 \nl
% R12 & $10^{12}$ & 10  & 1000 & 0.36 & 0.20 & 0.037  & 0.04    & 2.52    & 1.24  \nl
% S12 & $10^{12}$ & 0.5 & 5000 & 0.28 & 0.25 & 0.023  & 0.095   & 2.82    & 3.08  \nl
\enddata
%%\tablenotetext{a}{}
\tablecomments{
The model parameters, $B_0$, $\Omega_0$, and $R_0$, are shown in units of 
G, $\rm rad\,s^{-1}$, and km, respectively.
The calculations are stopped at time $t_f$.
The jets pass through a point 1000km from the central black hole
at time $t_{\rm jet}$.
The mass ejected through the jets, $M_{\rm ej}$, is in units of $M_\odot$.
The magnetic, kinetic, and internal energies
carried away by the jets, 
$(E_m)_{\rm ej}$, $(E_k)_{\rm ej}$, and $(E_i)_{\rm ej}$, 
are expressed in units of $10^{50} \rm erg$.}
\end{deluxetable}

\end{document}